\documentclass[desactivate]{aa}

\DeclareRobustCommand{\rchi}{{\mathpalette\irchi\relax}}
\newcommand{\irchi}[2]{\raisebox{\depth}{$#1\chi$}} 

\newcommand{\rlight}{r_\textrm{L}}

\usepackage{amsmath}
\usepackage{graphicx}
\usepackage{txfonts}
\usepackage{url}
\usepackage{hyperref}
\usepackage{color}

\begin{document} 

\title{Multi-wavelength emission modelling of PSR~J0437$-$4715}


\author{J. P\'etri\inst{1}
         \and P. Stammler\inst{2}
         \and L. Guillemot\inst{3,4}
         \and S. Guillot\inst{2}
         \and D. Gonz\'alez-Caniulef\inst{2}
         \and F. Jankowski \inst{3}
         \and N. Webb\inst{2}
          }

\institute{Universit\'e de Strasbourg, CNRS, Observatoire astronomique de Strasbourg, UMR 7550, F-67000 Strasbourg, France.\\
\email{jerome.petri@astro.unistra.fr}   
\and 
IRAP, CNRS, 9 avenue du Colonel Roche, BP 44346, F-31028 Toulouse Cedex 4, France
\and
LPC2E, OSUC, Univ Orl\'eans, CNRS, CNES, Observatoire de Paris, F-45071 Orl\'eans, France
\and 
ORN, Observatoire de Paris, Universit\'e PSL, Univ Orl\'eans, CNRS, 18330 Nan\c{c}ay, France
}

\date{Received ; accepted }

 
\abstract
{The diversity of pulsar light-curves and radio polarisation properties originates in the structure of the magnetic field close to the stellar surface. For millisecond pulsars, this complexity is particularly puzzling. Fortunately, some means exist to uncover the magnetic field topology which indeed impacts the emission within the magnetosphere but also on the surface through its hot spot thermal radiation.}
{We aim at deducing a plausible magnetic field geometry for the millisecond pulsar J0437$-$4715 by using combined information from the soft X-ray hot spot geometry deduced from NICER observations by pulse profile modelling and from radio and $\gamma$-ray pulse profile fitting. We also check the consistency between the geometry obtained and the radio polarisation data.}
{Our $\gamma$-ray light-curve shapes rely on the striped wind model, whereas the radio polarisation fits rely on the rotating vector model. The magnetosphere structure is obtained from dipolar force-free magnetosphere simulations.}
{We demonstrate that a slightly off-centred dipole augmented by a small scale dipole located on one polar cap explains simultaneously the shape of the hot spot and the radio and $\gamma$-ray data with a magnetic obliquity of $\alpha \approx (42\pm5) \degr$ and a line-of-sight inclination angle of $\zeta \approx (136 \pm5) \degr$.}
{Our simple dipole model reproduces all the radio and $\gamma$-ray characteristics of PSR~J0437$-$4715, including its radio polarisation data. It shows that the radio emission could be produced in regions where the magnetic field is mainly of dipolar nature.}


\maketitle

%

\section{Introduction}

Millisecond pulsars (MSPs) are fast rotating neutron stars spun up during their accretion phase in a binary system. This mechanism leads to rotational periods as fast as several milliseconds. They are old pulsars, aged billions of years, with weak surface magnetic field strengths of the order of $10^3$~T$-10^5$~T ($10^7$~G$-10^9$~G). 
Deciphering their magnetic field structure is a difficult task due to the small size of the light-cylinder and due to the strong impact of near surface multipolar or small scale magnetic fields. Nevertheless, attempts have been made to extract this field geometry by combining radio, soft X-ray and $\gamma$-ray emission in several MSPs observed with the Neutron star Interior Composition ExploreR (NICER), such as PSR~J0030+0451 \citep{petri_constraining_2023-2} or PSR~J0740+6620 \citep{petri_double_2025-2}. Determining the shape of the hot spots located at the neutron star surface offers a valuable insight into the near surface magnetic field configuration and could serve as an input for global magnetospheric simulations. This task was undertaken by using NICER data to analyse four MSPs like PSR~J0030+0451 \citep{riley_nicer_2019, miller_psr_2019, vinciguerra_updated_2024}, PSR~J0740+6620 \citep{salmi_radius_2022, miller_radius_2021, dittmann_more_2024} and PSR~J1231$-$1411 \citep{salmi_nicer_2024}, see also \cite{bogdanov_constraining_2019-1} for earlier results. More recently \cite{choudhury_nicer_2024} analysed PSR~J0437$-$4715 and found two non-antipodal hot spots whereas \cite{mauviard_nicer_2025} studied PSR~J0614--3329 and found two hot spots, one close to the pole and the other close to the equator. PSR~J0437$-$4715 is the nearest and brightest MSP known so far with soft X-ray emission and the target of interest in the present study. 

The discovery of the X-ray emission of PSR~J0437--4715 dates back to observations with the ROSAT (ROntgenSATellit) telescope \citep{becker_detection_1993}. The thermal nature of this emission was used for detailed modelling of its spectrum and of the pulsed profile resulting from the neutron star rotation \citep{zavlin_x-radiation_2002, zavlin_xmm-newton_2006}.  \cite{bogdanov_nearest_2013} performed the first phase-resolved spectroscopic study of this pulsar, and provided some constraints on its compactness.  Later, a broad-band X-ray analysis demonstrated the presence of three thermal components (the warm surface and two hot spots), characterised the non-thermal spectral component and showed pulsations up to $\sim 15$~keV with data from the Nuclear Spectroscopic Telescope Array (NuSTAR) \citep{guillot_nustar_2016}. The warm surface, at $T_{\!\rm eff}\sim 2\times 10^{5}$~K, was further studied by combining the X-ray data and far-ultraviolet observations with the Hubble Space Telescope \citep{gonzalez-caniulef_neutron_2019}.  In the radio, this pulsar has been known since the 1990s \citep{johnston_discovery_1993}, and its polarised radio emission was studied in \cite{navarro_mean_1997}. Single radio pulse analysis was detailed in \cite{oslowski_timing_2014}. 

From a theoretical point of view, several groups tackled the problem of connecting the hot spot geometry to the external magnetic field computed from the magnetosphere, including, for instance, a quadrupole component as in \cite{lockhart_x-ray_2019}. Multipolar components were also suspected in PSR~J0030+4051 \citep{kalapotharakos_multipolar_2021}. \cite{huang_physics-motivated_2025}, based on a force-free magnetosphere of an off-centred magnetic dipole, deduced the polar cap temperature distribution and predicted the observed X-ray light curves. \cite{chen_numerical_2020} did a similar analysis for PSR~J0030+0451 using a dipole \& quadrupole configuration. \cite{dyks_radio_2019} explained the complexity of the radio polarisation data in terms of a coherent superposition of orthogonal modes. Finally, \cite{sur_radio_2024} performed force-free simulations with multipoles and applied it to MSPs. One hot spot of PSR~J0437$-$4715 as found by \cite{choudhury_nicer_2024} shows an annular shape that is interpreted as the presence of a quadrudipole field, i.e. a combination of a dipole and a quadrupole \citep{gralla_inclined_2017}. More quantitatively, \cite{carrasco_relativistic_2023} performed general-relativistic force-free (GRFFE) simulations of neutron star magnetospheres to model the soft X-ray pulse profile of several NICER pulsars, including PSR~J0437$-$4715. Their work is based on the electric current circulating within the magnetosphere in the force-free regime, leading to non-standard emission regions (different from the almost circular polar cap regions).

PSR~J0437$-$4715 is the fourth pulsar with pulse-profile modelling analyses published by the NICER team. It is detected in radio and $\gamma$-rays with good quality radio polarisation data, as summarised in Sec.~\ref{sec:observations}. Based on these multi-wavelength observations, we infer the magnetic field geometry of the radio emission site, as explained in Sec.~\ref{sec:magnetic_field}. Conclusions are drawn in Sec.~\ref{sec:conclusion}.


\section{PSR~J0437$-$4715 multi-wavelength observations\label{sec:observations}}

\subsection{Observational facts}
PSR~J0437$-$4715 is an MSP spinning at a period of $P=5.75$~ms and orbiting in a binary system with a white dwarf companion at a period of $P_{\rm orb} = 5.74$~day. Its orbital inclination angle, derived from Shapiro delay measurements, is accurately determined to be $i= 137.506\degr \pm 0.016\degr$ \citep{reardon_neutron_2024} and its mass found to be ${M = 1.418\,M_\odot \pm 0.044 \, M_\odot}$. Pulse profile modelling performed by \cite{choudhury_nicer_2024} tends to a radius of $R = 11.36^{+0.96}_{-0.63}$~km. This leads to a ratio between the neutron star radius and its light-cylinder radius of $a = R/\rlight \approx 0.0414$ where $\rlight = c\,P/2\pi \approx 274$~km and a polar cap angular size of $\theta_{\rm pc} \approx \arcsin \sqrt{a} \approx 0.205$. 

\subsection{Radio and $\gamma$-rays observations}
\begin{figure}[h]
	\centering
	\includegraphics[width=\linewidth]{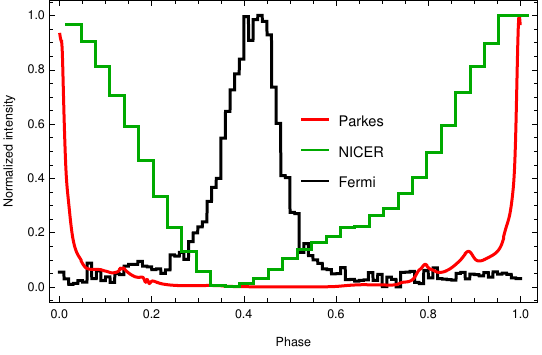}
	\caption{Multi-wavelength pulse profile of PSR~J0437$-$4715 as observed in radio by Parkes at $1.4$~GHz (red line), by NICER in soft X-rays (0.3-3.0~keV, green line) and in $\gamma$-rays by Fermi/LAT (black line). Note that the zero line of the NICER data is arbitrary, since technically, the pulsed emission lies on top of an unpulsed pulsar emission + background emission (all origins combined). Note that this is also true for the radio data, for which the profile baseline has been subtracted to lie at zero. }
	\label{fig:j0437-4715multilambda}
\end{figure}

PSR~J0437$-$4715 was among the first radio MSPs detected in $\gamma$-rays with the Fermi/LAT \citep{abdo_population_2009}. Updated $\gamma$-ray emission properties were more recently published in the third $\gamma$-ray pulsar catalogue (3PC) \citep{smith_third_2023}. Its multi-wavelength pulse profile, taken from 3PC for the radio and $\gamma$-ray part and from NICER for the X-ray part (see next subsection), is shown in Fig.~\ref{fig:j0437-4715multilambda}.
It shows a single wide radio pulse spanning almost the full period with a duty cycle of 80\%. The single $\gamma$-ray peak is separated from this radio peak by approximately $0.4$ in phase, whereas the soft X-ray peak is almost aligned with the radio pulse peak.

Good radio polarisation data have also been published by \cite{dai_study_2015} and are available on the EPN database. The polarisation position angle (PPA) is complex and does not reflect the rotating vector model (RVM) expectation, probably due to multipolar components or to emission height variation, twisted magnetic field lines and orthogonal polarisation modes.

\subsection{NICER data}

The NICER X-ray pulse profile of PSR~J0437$-$4715 has been generated following the standard data reduction procedure\footnote{\url{https://heasarc.gsfc.nasa.gov/docs/nicer/analysis_threads/nicerl2/}} on the X-ray Timing Instrument (XTI) observations from 2017-07-06:T14:34:12 to 2023-05-21:T22:22:27 (ObsIDs: 0060010101-6060010721). Because of an optical light leak\footnote{\url{https://heasarc.gsfc.nasa.gov/docs/nicer/analysis_threads/light-leak-overview/}} resulting from NICER thermal film damage on 2023-05-22, observations since this date have been ignored. 

Moreover, a handful of observations presented time stamp anomalies on the detectors of the Measurement Power Unit~1, because of a single-event upset\footnote{\url{https://heasarc.gsfc.nasa.gov/docs/nicer/data_analysis/nicer_analysis_tips.html}} affecting the XTI (affected ObsIDs: 2060010407-2060010412) when NICER was passing through the South Atlantic Anomaly. Those events have been excluded as well. The remaining data have been calibrated and filtered with the \textit{nicerl2} tool from \texttt{NICERDAS v13}, provided with \texttt{HEASOFT v6.34}, and the \texttt{CALDB} version \texttt{20240206}. Hereafter, in order to minimise background contributions, we applied further filtering cuts using the \texttt{NicerSoft}\footnote{\url{https://github.com/paulray/NICERsoft}} package, selecting events with i) a cutoff rigidity $COR_{-}SAX > 1.5\,\mathrm{GeV\,c}^{-1}$, ii) a planetary K index $K_P < 5$, iii) Sun angles $> 60 \degr$, iv) overshoot rates $< 1.5\,\mathrm{counts\,s}^{-1}$ and v) undershoot rates $< 200\,\mathrm{counts\,s}^{-1}$. A last filtering cut was performed by excluding any good time interval with more than 4 counts\,s$^{-1}$ in the 3--10 keV range, and obtained a final good exposure time of $2.56\,\mathrm{Ms}$. We ultimately phase-folded the merged event list with the \texttt{PINT}\footnote{\url{https://github.com/nanograv/PINT}} \textit{photonphase} tool, generating a pulse profile in soft X-ray energies ($0.3-3.0$~keV), based on the ephemeris from the Fermi/LAT 3PC timing solution \citep{smith_third_2023}. 

\section{Magnetic field determination\label{sec:magnetic_field}}

As in our previous studies, we start by exploring the radio and $\gamma$-ray light-curves to constrain the large scale geometry of the dipole magnetic field and the associated magnetic obliquity angle $\alpha$ and line of sight inclination angle $\zeta$. In a second step, we check consistency of the hot spot location and size with the published NICER results. Finally, we extract additional information from the radio polarisation data.

\subsection{Dipole geometry from radio and $\gamma$-ray modelling}

\begin{figure}[h]
	\centering
	\includegraphics[width=\linewidth]{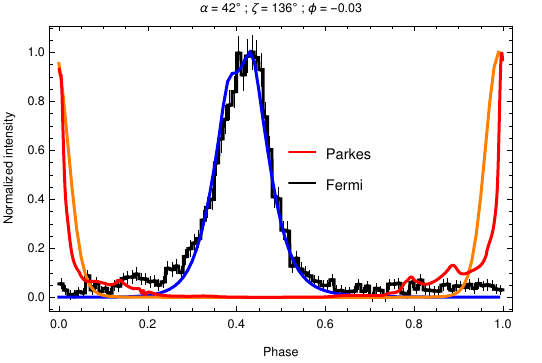}
	\caption{Example of a good fit of the $\gamma$-ray pulse profile ($\geq$ 0.1 GeV). The radio profile is shown in red, our model is displayed in orange, the $\gamma$-ray light-curve in black and its fit in blue.}
	\label{fig:gamma}
\end{figure}

The presence of a single $\gamma$-ray peak as reported in 3PC \citep{smith_third_2023} suggests that the angles $\alpha$ and $\zeta$ are roughly given by ${\zeta \approx \alpha \approx 45 \degr}$ or by their supplementary angles ${\pi-\zeta \approx \pi-\alpha \approx 135 \degr}$. However, in order to better constrain these angles, we fitted the $\gamma$-ray light-curve shown in Fig.~\ref{fig:j0437-4715multilambda}, using the same model as in \cite{petri_constraining_2023-2}, where a $\gamma$-ray atlas has been used. A good fit to this $\gamma$-ray light-curve is given by $\alpha=(42\pm5)\degr$ and $\zeta=(136\pm5)\degr$, as shown in Fig.~\ref{fig:gamma}. In order to accurately fit the $\gamma$-ray peak time lag with respect to the radio peak, we added a third parameter quantifying an additional phase shift not predicted by our model and denoted by~$\phi$ as done in several of our previous works. This additional parameter corrects for the slight mismatch between the predicted $\gamma$-ray peak location at a phase slightly larger than the observed one at around $\varphi\approx 0.4$. Its value remains however small with $\phi \approx -0.03$ showing that our model reasonably well explains the radio time lag. In other words, the predicted light-curve intensity is $\mathbb{I}(\varphi)$ whereas to account for the observation, we need to shift it by a small lag $\phi$ such that the best fitted light-curve becomes $\mathbb{I}(\varphi-\phi)$. This first fit gives the overall description of the dipole field, dominant at large distances. 
The line of sight inclination angle $\zeta$ we found is equal to the orbital inclination angle $i$ within the uncertainties (actually we used increments of $1\degr$ for $\zeta$) suggesting that this binary system shows nearly perfect alignment between the pulsar rotation axis and the orbital angular momentum, $\zeta \approx i$.

\begin{figure}
	\centering
	\includegraphics[width=\linewidth]{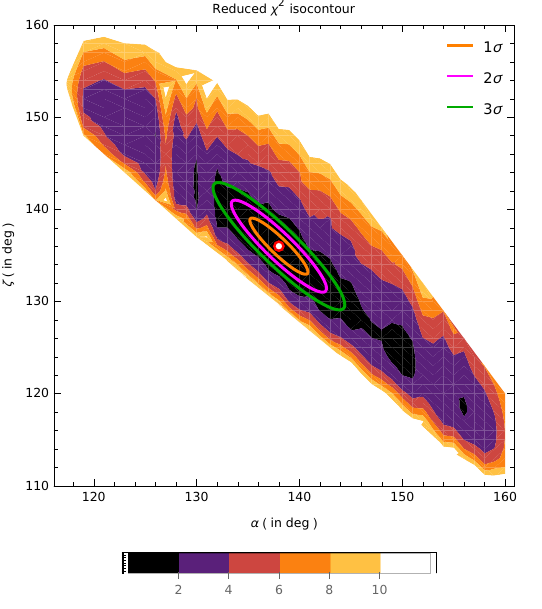}
	\caption{Color map showing the isocontours of the reduced $\rchi^2$ fit for the $\gamma$-ray light-curve for the angles $\alpha$ and $\zeta$. The $1\sigma$, $2\sigma$ and $3\sigma$ confidence intervals are also shown. The minimum is located at ${(\alpha,\zeta)=(138\degr,136\degr)}$ (corresponding to the mirror angle ${\alpha=42\degr}$) and depicted by a red circle.}
	\label{fig:carte_residu}
\end{figure}

The quality of the fit is checked by the value of the reduced $\rchi^2$ as shown in Fig.~\ref{fig:carte_residu}. The minimum value is around $1.3$ and the uncertainties in the angles $\alpha$ and $\zeta$ about $5\degr$. Note however, that because the $\gamma$-ray light-curve is invariant with respect to the change $\alpha \rightarrow \pi - \alpha$ and because we discuss later the radio polarisation fits, shown in black solid contours, we decided to plot the $\rchi^2$ values in the mirror symmetric region for $\alpha>90\degr$.

\subsection{Dipole geometry from X-ray pulsations}

We next switch to the X-ray data interpretation and summarise the findings of \cite{choudhury_nicer_2024} in Table~\ref{tab:NICER1} for their credible interval (CI) estimate and their maximum likelihood (ML) value. Uncertainties about the given angles are about $0.05$~rad, thus about $3\degr$. The hot spots are non-antipodal and could correspond to an off-centred dipole. The geometry of this plausible dipole with parameters $(\alpha, \beta, \delta, \epsilon$, see \citealt{petri_polarized_2017}, for the definition of these angles and $\epsilon$) is then deduced and shown on the right columns in the same Table~\ref{tab:NICER1}. We found a mildly off-centred structure with a displacement $d$ such that $\epsilon = d/R \approx 0.36-0.46$ and an obliquity about $\alpha=20\degr-26\degr$. This is however $15\degr$ to $20\degr$ less than the magnetic obliquity $\alpha$ found by $\gamma$-ray light-curve fitting, although the inclination angle is within $1\degr$ of the radio measurement \citep{reardon_neutron_2024} used in the NICER analyses. Nevertheless, the uncertainties in the location of the centre of the hot spot amounts to several degrees, allowing for some freedom and could increase the expected inclination of the equivalent magnetic dipolar axis to remain consistent with the $\gamma$-ray light-curve predictions.
\begin{table*}[h]
	\centering
\caption{Off-centred dipole geometry deduced from the polar cap location after the new joint NICER and XMM-Newton results of \cite{choudhury_nicer_2024}. The subscripts ${\rm p}$ and ${\rm s}$ stand for primary and secondary hot spot. See \cite{petri_polarized_2017} for the definition of the angles $(\alpha,\beta,\delta)$ and the displacement $\epsilon$.  \label{tab:NICER1}}
\begin{tabular}{lcccccccc|cccc}
	\hline
    & \multicolumn{8}{c}{From the literature} & \multicolumn{4}{|c}{From this work} \\
    \hline
	Hot spot & $\Theta_{\rm p}$ & $\phi_{\rm p}$ & $\xi_{\rm p}$ & $\Theta_{\rm c,s}$ & $\phi_{\rm s}$ & $\xi_{\rm c,s}$ & $\xi_{\rm c,s}/\xi_{\rm p}$ & $\zeta$ & $\alpha$ & $\beta$ & $\delta$ & $\epsilon$ \\ 
	Model & (rad) & (cycle) & (rad) & (rad) & (cycle) & (rad) &  &  &  &  &  & ($d/R$) \\
	\hline
	CI & 0.146 & 0.4429 & 0.433 & 2.307 & 0.4704 & 0.197 & 0.45 & 137.5\degr & 20\degr & 110\degr & 70\degr & 0.46 \\
	ML & 0.112 & 0.4381 & 0.561 & 2.318 & 0.466  & 0.177 & 0.31 & 137.5\degr & 26\degr & 116\degr & 64\degr & 0.36 \\
	\hline
\end{tabular}
\end{table*}

Moreover, taking into account the oblate shape and the associated gravitational field could have an impact on the above results. \cite{cadeau_light_2007} showed that for spin frequencies above 300~Hz, large deviations from the correct X-ray light curve are expected if a spherical star is used, see also \cite{morsink_oblate_2007} for the impact of oblateness. This should be accurately quantified, but is outside the scope of this work.

\begin{figure}[h]
	\centering
	\includegraphics[width=\linewidth]{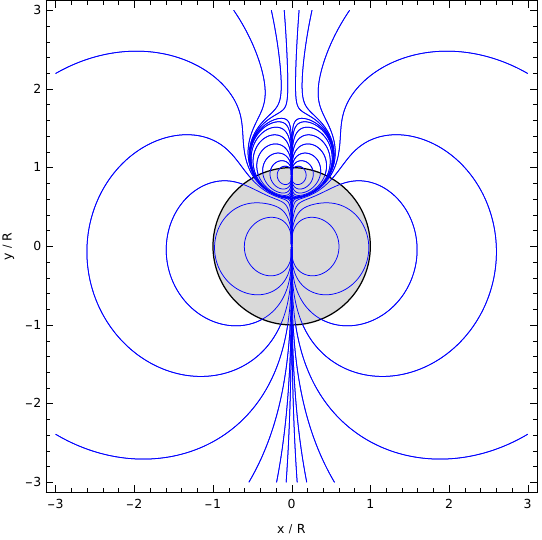}
	\caption{Example of a dipole producing an annular hot spot on one pole. Magnetic field lines are shown in blue and the neutron star surface as a solid black circle of normalised radius. The stellar interior is shown in light gray.}
	\label{fig:double_dipole}
\end{figure}

The annular shape of the northern hot spot found from the NICER data analyses \citep{choudhury_nicer_2024} could be produced by a small scale dipole underneath the surface, as shown in Fig.~\ref{fig:double_dipole}. This is an alternative to the dipole+quadrupole geometry presented in other works \citep{gralla_inclined_2017}. In order to recover the size and thickness of this annular ring, the strength and precise location of this small scale dipole must be adjusted. However, we do not dive into such details that would not inform us more about the multi-wavelength emission, as this pole is not detected in radio.

\subsection{Radio pulse profile}

\begin{figure}[h]
	\centering
	\includegraphics[width=\linewidth]{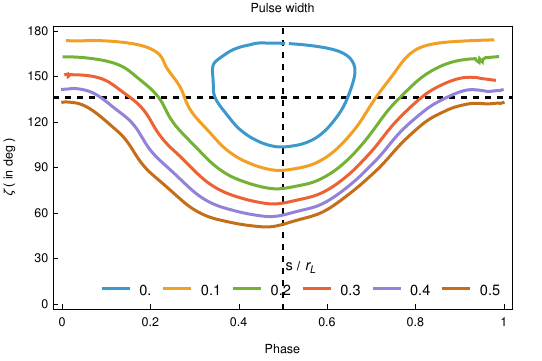}
	\caption{Projection of the emission cone rims onto the sky for different emission lengths $s$ in units of $\rlight$. The phase location of the magnetic axis at $0.5$ is shown as a vertical dashed line, whereas the line of sight at $\zeta=136\degr$ is shown as a horizontal dashed line.}
	\label{fig:largeur_pulse_a42}
\end{figure}

To go further in our understanding of the emission location and geometry, we investigate features of the radio pulse profile. Starting with the pulse width~$W$ related to the beam cone half-opening angle~$\rho$ by \citep{gil_geometry_1984}:
\begin{equation}\label{key}
	\cos \rho = \cos \alpha \cos \zeta + \sin \alpha \sin \zeta \cos \, (W/2), 
\end{equation}
we estimate the cone opening angle to be $\rho \approx 80\degr$, assuming $\alpha=42\degr$, $\zeta=136\degr$ and a pulse width of 80\% of the period thus $W=0.8\times2\pi$.
For a pure static dipole, this angle is related to the colatitude $\theta$ at which the radio beam is formed by
\begin{equation}
	\rho = \theta + \arctan (\tan \theta/2) \approx 3 \, \theta /2 \ .
\end{equation}
Therefore, $\theta \approx 53\degr$ and the emission height becomes $h/\rlight \approx \sin^2 \theta \approx 0.64$, thus, a substantial fraction of the light-cylinder radius. We can perform better by computing the separatrix surface deduced from the force-free model and projecting the tangent to the field lines onto the sky. The result is shown in Fig.~\ref{fig:largeur_pulse_a42} for $\alpha=42\degr$. The magnetic axis is located at phase $0.5$ and shown as a vertical dashed line, whereas the line of sight at $\zeta=136\degr$ is shown as a horizontal dashed line. Each coloured line corresponds to a constant length $s$ along a field line starting at the stellar surface and given in units of $\rlight$. Note that this length does not correspond to the usual emission altitude $h$ reported in other works, as field lines are curved and not straight lines in the radial direction. The arc length $s$ is measured in the neutron star corotating frame and starts with $s=0$ at the neutron star surface whereas $h$ starts at the star centre, thus $h=R$ corresponds to $s=0$. The curve asymmetry in Fig.~\ref{fig:largeur_pulse_a42} steams mainly from the electric current induced by the plasma flow and by the magnetic field sweep-back.

\begin{figure}[h]
	\centering
	\includegraphics[width=\linewidth]{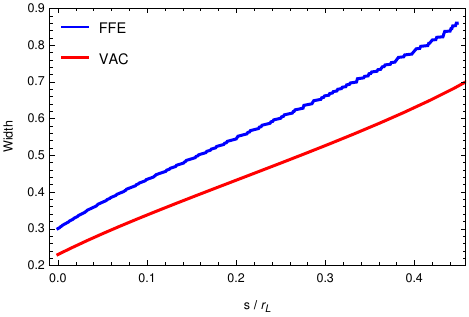}
	\caption{Width of the radio pulse profile depending on the emission height~$s$ above the stellar surface, along the magnetic field lines, in units of $\rlight$ for $\alpha=42\degr$ and $\zeta=136\degr$, in blue for force-free (FFE) and in red for vacuum (VAC).}
	\label{fig:widthvsheight}
\end{figure}

The evolution of the pulse width with respect to this curvilinear abscissa is shown in Fig.~\ref{fig:widthvsheight} in blue for the force-free (FFE) case and in red for the vacuum (VAC) case. It is a monotonically increasing function of the emission length~$s$ and a pulse width of 80\% with $W=0.8\times2\pi$ requires a length of about $s/\rlight=0.4$, thus a height of $h/\rlight\approx 0.5$ slightly less than the pure dipole estimates. Above $s/\rlight=0.5$, the pulse duty cycle becomes 100\% and is visible for the full period of the pulsar. Fig.~\ref{fig:largeur_pulse_a42} demonstrates that detecting pulsations for a significant fraction of the period, of almost 100\%, does not require an almost aligned rotator for these pulsar parameters, especially fast rotation.

\subsection{Radio polarisation}

In order to strengthen our estimate of the magnetic obliquity and line of sight inclination angles, we conclude this study by including radio polarisation data as observed for instance by \cite{oslowski_timing_2014} and \cite{dai_study_2015}. Multi-frequency polarisation pulse profiles are available at wavelengths of 10~cm ($3.0$~GHz), 20~cm ($1.5$~GHz) and 50~cm ($0.6$~GHz) with the associated PPA from the European Pulsar Network (EPN) database website. At 50~cm, the evolution is rather monotonic, with only two or three orthogonal polarisation mode (OPM) switching events. At 20~cm, the smoothness becomes less obvious with stronger OPM and at 10~cm the PPA becomes much more erratic. This shift from a well-behaved PPA at low frequency to a much less well-structured PPA at high frequency hints at a decrease in the emission height when increasing the radiation frequency, in agreement with the radius-to-frequency mapping picture. Lower altitude means stronger multipolar components and therefore possibly significant deviation from the RVM expectations.
Nevertheless, let us fit the rotating vector model of \cite{radhakrishnan_magnetic_1969}, who gave the time evolution of the linear polarisation angle $\psi$ as
\begin{equation}
	\label{eq:RVM}
	\tan (\psi - \psi_0) = \frac{\sin \alpha \, \sin (\varphi-\varphi_0)}{\cos \alpha \, \sin \zeta - \sin \alpha \, \cos \zeta \, \cos (\varphi-\varphi_0)} \ ,
\end{equation}
where $\psi_0$ corresponds to the position angle of the spin axis on the plane of the sky and $\varphi$ is the rotational phase ($\varphi_0$ is a reference phase corresponding to the inflexion point, obtained by fitting the polarisation to the RVM expectations). 

\begin{figure}[h]
	\centering
	\includegraphics[width=\linewidth]{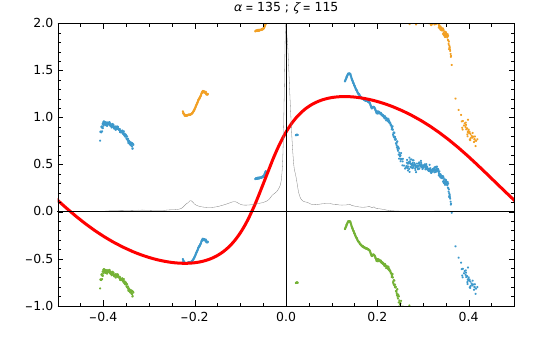}
	\includegraphics[width=\linewidth]{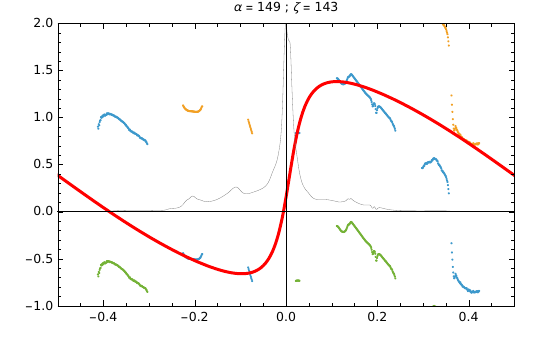}
	\includegraphics[width=\linewidth]{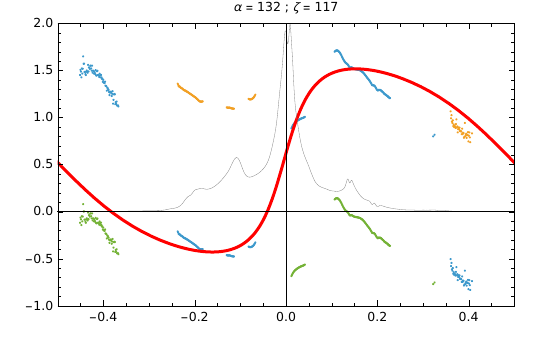}
	\caption{PPA as a function of rotational phase for PSR~J0437$-$4715 as measured with the Parkes telescope at 10~cm, 20~cm and 50~cm (top to bottom panels). Bins with polarisation degree $\Pi_i<30\%$ have been removed. The best values for the PPA fits from the RVM are reported in Table~\ref{tab:RVMfit}. The radio pulse profile is shown in gray for better identification of the pulse. The RVM fit is shown in red solid line.}
	\label{fig:polarisation}
\end{figure}

Neglecting the presence of OPM would lead to bad fits to the PPA. Identifying the OPM jumps or simply removing them "by eye" is not satisfactory. We therefore designed a fitting method that automatically incorporates variations of $\pm90\degr$ into the fitting scheme. To achieve our goal, we introduce a distance $d(y_i, m_i)$ between the observations $y_i$ and model $m_i$ for each bin~$i$ such that 
\begin{subequations}\label{key}
	\begin{align}
	d(y_i, m_i) & =\text{min}(\Delta_i ,\pi-\Delta_i ,\Delta_i^+, \pi-\Delta_i^+) \\
	\Delta_i & = |y_i - m_i| \\
	\Delta_i^+ & = |y_i - (m_i+\pi/2)| \ .
	\end{align}
\end{subequations}
This distance takes into account the periodicity of $180\degr$ ($\pi$~rad) of the PPA as well as OPM switching for each bin. The values $y_i$ and $m_i$ are forced to stay in the interval $[0,\pi]$ during the computation. 
We then minimise the following cost function, a kind of $\chi^2$ function defined by
\begin{equation}
	\label{eq:chi2radio}
	\chi^2 = \sum_{i} \frac{d(y_i, m_i)^2}{\sigma_i^2} \ .
\end{equation}
It automatically identifies the number of OMP jumps and their phase locations. $\sigma_i$ represents the uncertainties in the polarisation angle in each bin~$i$. Actually, because PSR~J0437-4715 is very bright in radio, these uncertainties are tiny in general, less than $1\degr$ even less than $0.1\degr$ in some phase bins. We therefore do not expect to obtain good reduced $\chi^2$ numbers close to unity because of the OPM and small structures in the PPA. Moreover, since a low degree of linear polarisation $\Pi_i = L_i/I_i$ ($I_i$ being the total intensity Stokes parameter and $L_i = \sqrt{U_i^2+Q_i^2}$ the linear polarisation intensity obtained from $Q_i$ and $U_i$ Stokes parameters in each bin $i$) makes it difficult to distinguish between the two orthogonal modes, we impose a threshold on $\Pi_i$ to improve the distinction between the OPM. We also modify the uncertainties $\sigma_i$ to weighted uncertainties related to degree of linear polarisation $\Pi_i$ such that $\sigma_i$ is replaced by $\sigma_i'^{-1} = I_i \, \Pi_i = L_i$, thus focusing on bright and highly linearly polarised bin phases. 
We typically set $\Pi_i = 20\%-50\%$, removing data points below this threshold. We also enforce a threshold for the inflexion point such that $\sin\alpha \geq 2 \, \sin (\alpha - \zeta)>0$ in order to constrain the derivative of the PPA to have the right trend with phase. This is particularly useful at high polarisation $\Pi_i$ threshold where data points are almost absent in the middle of the pulse profile. 
The resulting fits, using the modified $\chi^2$ approach \eqref{eq:chi2radio}, are therefore performed only on a subset of the total observational data, as shown in Fig.~\ref{fig:polarisation} where bins with polarisation degree $\Pi_i<30\%$ have been removed. The orange and green data points show the PPA when shifted by $\pm90\degr$ to simulate orthogonal polarisation modes. The RVM shows a reasonable fit to the data with $\alpha_{\rm rvm}=132\degr-149\degr$ and ${\zeta_{\rm rvm}=114\degr-143\degr}$ at different wavelengths of 10~cm, 20~cm and 50~cm. The exact values of the fitting for the different wavelengths are reported in Table~\ref{tab:RVMfit}. 

Note that $\alpha_{\rm rvm}$ is the supplementary to $\alpha$ because we focus on the South Pole for polarisation purposes. Following the $\gamma$-ray light-curve fitting value, we should get ${\alpha_{\rm rvm} \approx \pi - \alpha \approx 180\degr - 42\degr = 138 \degr}$. Our values in Table~\ref{tab:RVMfit} are consistent with this guess within the uncertainties. 
\begin{table}[h]
	\centering
    \caption{Radio pulse width, at 10\% maximum intensity, obliquity $\alpha$, line of sight $\zeta$, phase shift $\varphi_0$ and offset $\psi_0$ obtained from RVM fits at different wavelengths and with minimum polarisation bins ${\Pi_i>30\%}$. The to last columns correspond to the combined radio+$\gamma$ fit.\label{tab:RVMfit}}
	\begin{tabular}{cccccc}
		\hline
		Wavelength & width & \multicolumn{2}{c}{Radio only} & \multicolumn{2}{c}{Radio+$\gamma$} \\
		& $w_{10}$ & $\alpha$ & $\zeta$ & $\alpha$ & $\zeta$ \\
		\hline
		$10$~cm & $0.777$ & $135\degr$ & $115\degr$ & $145\degr$ & $129\degr$ \\
		$20$~cm & $0.835$ & $149\degr$ & $143\degr$ & $140\degr$ & $131\degr$ \\
		$50$~cm & $0.877$ & $132\degr$ & $117\degr$ & $140\degr$ & $135\degr$ \\
		\hline
	\end{tabular}
\end{table}




\subsection{Combined $\gamma$+radio fit}

\begin{figure}[h]
	\centering
	\includegraphics[width=0.8\linewidth]{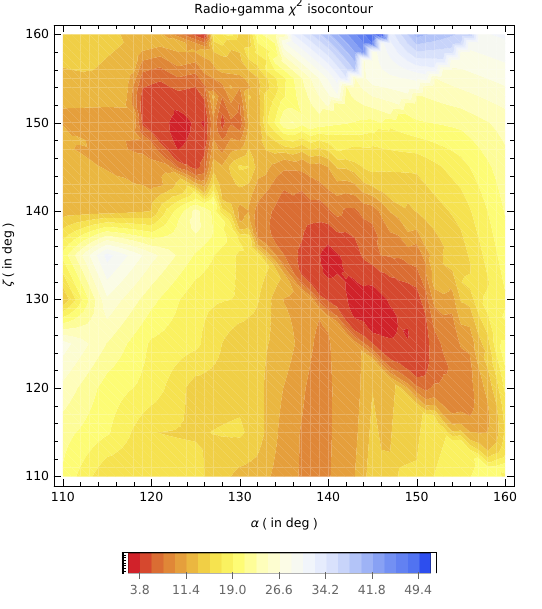}
	\includegraphics[width=0.8\linewidth]{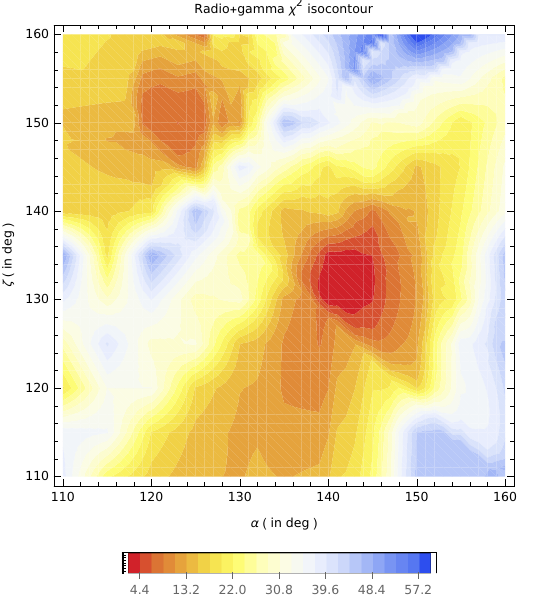}
	\includegraphics[width=0.8\linewidth]{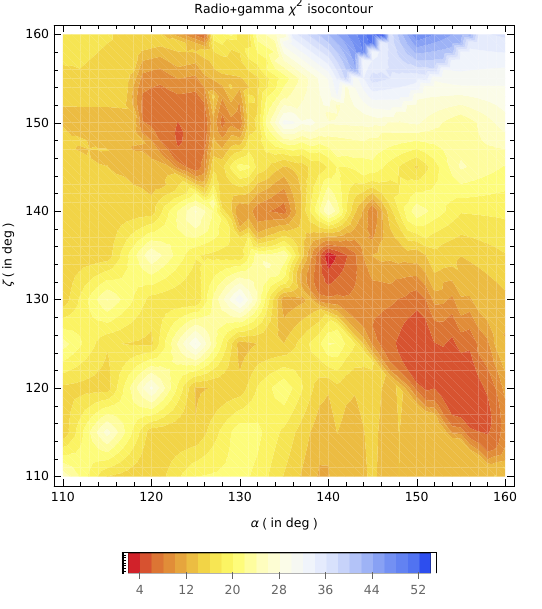}
	\caption{Combined radio and $\gamma$-ray $\chi^2$ maps of Eq.\eqref{eq:chi2_combine} at the three wavelengths at $10$~cm, $20$~cm and $50$~cm, from top to bottom.}
	\label{fig:chi2combine_50cm_gamma}
\end{figure}

\begin{figure}[h]
	\centering
	\includegraphics[width=\linewidth]{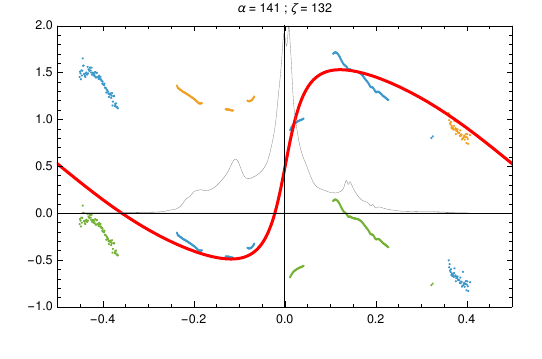}
	\caption{Best joined radio/$\gamma$-ray fit with $(\alpha,\zeta)=(141\degr,132\degr)$ to be compared with the bottom panel of Fig.~\ref{fig:polarisation} where bins with linear polarisation degree less than $\Pi_i<30\%$ have been removed.}
	\label{fig:polarisation_50cm_gamma}
\end{figure}

In a last step, we combine our $\gamma$ and radio fits in a weighted total $\chi^2$ value given by
\begin{equation}\label{eq:chi2_combine}
	\chi^2 = w_\gamma \, \chi^2_\gamma + w_r \, \chi^2_r \ .
\end{equation}
Because $\chi^2_\gamma$ and $\chi^2_r$ have usually very different values, we adjust the weights $w_\gamma$ and $w_r$ such that both $\chi^2$ contribute equally at their respective minimum. In other words, we choose $w_\gamma = 1/\textrm{min}(\chi^2_\gamma)$ and $w_r = 1/\textrm{min}(\chi^2_r)$ to avoid dominance of one wavelength against the other. A summary of the joint best fit is given in the last two columns of Table~\ref{tab:RVMfit} and the combined $\chi^2$ maps are shown in Fig.~\ref{fig:chi2combine_50cm_gamma} for the three wavelengths at $10$~cm, $20$~cm and $50$~cm. 
If the $\gamma$-ray fit is combined with the radio polarisation fit, for instance at the $50$~cm wavelength, we obtain the best compromise between the radio polarisation and the $\gamma$-rays by averaging the obliquity and line of sight angles such that $(\alpha,\zeta) \approx (141\degr,132\degr)$. 
The impact of this geometry onto the PPA evolution is plotted in Fig.~\ref{fig:polarisation_50cm_gamma} at 50~cm. 


\section{Conclusions\label{sec:conclusion}}

Although MSP emission is usually thought to be difficult to model with a dipolar magnetic field because of the impact of multipolar components close to the stellar surface, we showed that a modified dipole confidently reproduces a wealth of multi-wavelength properties of PSR~J0437$-$4715. Independent determinations of the viewing angle agree to very good accuracy with $\zeta=(136\pm5)\degr$. Even the radio polarisation data are satisfactorily fitted with the RVM and a large duty cycle in radio does not imply an almost aligned rotator. All these results give us confidence in the reliability of $\gamma$-ray light-curve fitting within the striped wind framework, even for MSPs. This work supports once more the idea that two off-centred dipoles can satisfactorily reproduce the multi-wavelength feature of NICER pulsars, as well as the hot spot geometry \citep{petri_constraining_2023-2,petri_double_2025-2}. Several other NICER pulsars like PSR~J1231$-$1411 and PSR~J0614$-$3329 will be investigated with the method presented in this work in order to improve our understanding of MSP magnetic field structure and geometry.

To constrain the radio part of the spectrum even more firmly, a single pulse study would greatly benefit the radio polarisation fit, as suggested by \cite{mitra_evidence_2023}. This study should keep only those that are highly linearly polarised. PSR J0437-4715 is a very bright radio pulsar, so that single-pulse studies should be possible, emphasizing again that a joint radio and $\gamma$ fit offers a very promising avenue to better understand MSP radio emission mechanisms.

\begin{acknowledgements}
We are grateful to the referee for helpful comments and suggestions. This work has been supported by the ANR (Agence Nationale de la Recherche) grant number {ANR-20-CE31-0010}. SG and PS acknowledge the support of the CNES. Part of this research has made use of the EPN Database of Pulsar Profiles maintained by the University of Manchester, available at: http://www.jodrellbank.manchester.ac.uk/research/pulsar/Resources/epn/.
\end{acknowledgements}


\begin{thebibliography}{40}
\expandafter\ifx\csname natexlab\endcsname\relax\def\natexlab#1{#1}\fi

\bibitem[{Abdo {et~al.}(2009)Abdo, Ackermann, Ajello, Atwood, Axelsson,
  Baldini, Ballet, Barbiellini, Baring, Bastieri, Baughman, Bechtol,
  Bellazzini, Berenji, Bignami, Blandford, Bloom, Bonamente, Borgland, Bregeon,
  Brez, Brigida, Bruel, Burnett, Caliandro, Cameron, Camilo, Caraveo, Carlson,
  Casandjian, Cecchi, Çelik, Charles, Chekhtman, Cheung, Chiang, Ciprini,
  Claus, Cognard, Cohen-Tanugi, Cominsky, Conrad, Corbet, Cutini, Dermer,
  Desvignes, de~Angelis, de~Luca, de~Palma, Digel, Dormody, do~Couto~e Silva,
  Drell, Dubois, Dumora, Edmonds, Farnier, Favuzzi, Fegan, Focke, Frailis,
  Freire, Fukazawa, Funk, Fusco, Gargano, Gasparrini, Gehrels, Germani,
  Giebels, Giglietto, Giordano, Glanzman, Godfrey, Grenier, Grondin, Grove,
  Guillemot, Guiriec, Hanabata, Harding, Hayashida, Hays, Hobbs, Hughes,
  Jóhannesson, Johnson, Johnson, Johnson, Johnson, Johnston, Kamae, Katagiri,
  Kataoka, Kawai, Kerr, Knödlseder, Kocian, Kramer, Kuss, Lande, Latronico,
  Lemoine-Goumard, Longo, Loparco, Lott, Lovellette, Lubrano, Madejski, Makeev,
  Manchester, Marelli, Mazziotta, McConville, McEnery, McLaughlin, Meurer,
  Michelson, Mitthumsiri, Mizuno, Moiseev, Monte, Monzani, Morselli,
  Moskalenko, Murgia, Nolan, Norris, Nuss, Ohsugi, Omodei, Orlando, Ormes,
  Paneque, Panetta, Parent, Pelassa, Pepe, Pesce-Rollins, Piron, Porter,
  Rainò, Rando, Ransom, Ray, Razzano, Rea, Reimer, Reimer, Reposeur, Ritz,
  Rochester, Rodriguez, Romani, Roth, Ryde, Sadrozinski, Sanchez, Sander,
  Saz~Parkinson, Scargle, Schalk, Sgrò, Siskind, Smith, Smith, Spandre,
  Spinelli, Stappers, Starck, Striani, Strickman, Suson, Tajima, Takahashi,
  Tanaka, Thayer, Thayer, Theureau, Thompson, Thorsett, Tibaldo, Torres, Tosti,
  Tramacere, Uchiyama, Usher, Van~Etten, Vasileiou, Venter, Vilchez, Vitale,
  Waite, Wallace, Wang, Watters, Webb, Weltevrede, Winer, \&
  Wood}]{abdo_population_2009}
Abdo, A.~A., Ackermann, M., Ajello, M., {et~al.} 2009, Science, 325, 848, aDS
  Bibcode: 2009Sci...325..848A

\bibitem[{Becker \& Trümper(1993)}]{becker_detection_1993}
Becker, W. \& Trümper, J. 1993, Nature, 365, 528, aDS Bibcode:
  1993Natur.365..528B

\bibitem[{Bogdanov(2013)}]{bogdanov_nearest_2013}
Bogdanov, S. 2013, ApJ, 762, 96

\bibitem[{Bogdanov {et~al.}(2019)Bogdanov, Guillot, Ray, Wolff, Chakrabarty,
  Ho, Kerr, Lamb, Lommen, Ludlam, Milburn, Montano, Miller, Bauböck, Özel,
  Psaltis, Remillard, Riley, Steiner, Strohmayer, Watts, Wood, Zeldes, Enoto,
  Okajima, Kellogg, Baker, Markwardt, Arzoumanian, \&
  Gendreau}]{bogdanov_constraining_2019-1}
Bogdanov, S., Guillot, S., Ray, P.~S., {et~al.} 2019, ApJ, 887, L25, aDS
  Bibcode: 2019ApJ...887L..25B

\bibitem[{Cadeau {et~al.}(2007)Cadeau, Morsink, Leahy, \&
  Campbell}]{cadeau_light_2007}
Cadeau, C., Morsink, S.~M., Leahy, D., \& Campbell, S.~S. 2007, ApJ, 654, 458

\bibitem[{Carrasco {et~al.}(2023)Carrasco, Pelle, Reula, Viganò, \&
  Palenzuela}]{carrasco_relativistic_2023}
Carrasco, F., Pelle, J., Reula, O., Viganò, D., \& Palenzuela, C. 2023, MNRAS,
  520, 3151

\bibitem[{Chen {et~al.}(2020)Chen, Yuan, \& Vasilopoulos}]{chen_numerical_2020}
Chen, A.~Y., Yuan, Y., \& Vasilopoulos, G. 2020, ApJL, 893, L38

\bibitem[{Choudhury {et~al.}(2024)Choudhury, Salmi, Vinciguerra, Riley, Kini,
  Watts, Dorsman, Bogdanov, Guillot, Ray, Reardon, Remillard, Bilous,
  Huppenkothen, Lattimer, Rutherford, Arzoumanian, Gendreau, Morsink, \&
  Ho}]{choudhury_nicer_2024}
Choudhury, D., Salmi, T., Vinciguerra, S., {et~al.} 2024, ApJL, 971, L20

\bibitem[{Dai {et~al.}(2015)Dai, Hobbs, Manchester, Kerr, Shannon, van Straten,
  Mata, Bailes, Bhat, Burke-Spolaor, Coles, Johnston, Keith, Levin, Osłowski,
  Reardon, Ravi, Sarkissian, Tiburzi, Toomey, Wang, Wang, Wen, Xu, Yan, \&
  Zhu}]{dai_study_2015}
Dai, S., Hobbs, G., Manchester, R.~N., {et~al.} 2015, MNRAS, 449, 3223

\bibitem[{Dittmann {et~al.}(2024)Dittmann, Miller, Lamb, Holt, Chirenti, Wolff,
  Bogdanov, Guillot, Ho, Morsink, Arzoumanian, \&
  Gendreau}]{dittmann_more_2024}
Dittmann, A.~J., Miller, M.~C., Lamb, F.~K., {et~al.} 2024, ApJ, 974, 295, aDS
  Bibcode: 2024ApJ...974..295D

\bibitem[{Dyks(2019)}]{dyks_radio_2019}
Dyks, J. 2019, MNRAS, 488, 2018

\bibitem[{Gil {et~al.}(1984)Gil, Gronkowski, \& Rudnicki}]{gil_geometry_1984}
Gil, J., Gronkowski, P., \& Rudnicki, W. 1984, A\&A, 132, 312

\bibitem[{González-Caniulef {et~al.}(2019)González-Caniulef, Guillot, \&
  Reisenegger}]{gonzalez-caniulef_neutron_2019}
González-Caniulef, D., Guillot, S., \& Reisenegger, A. 2019, MNRAS, 490, 5848

\bibitem[{Gralla {et~al.}(2017)Gralla, Lupsasca, \&
  Philippov}]{gralla_inclined_2017}
Gralla, S.~E., Lupsasca, A., \& Philippov, A. 2017, ApJ, 851, 137

\bibitem[{Guillot {et~al.}(2016)Guillot, Kaspi, Archibald, Bachetti, Flynn,
  Jankowski, Bailes, Boggs, Christensen, Craig, Hailey, Harrison, Stern, \&
  Zhang}]{guillot_nustar_2016}
Guillot, S., Kaspi, V.~M., Archibald, R.~F., {et~al.} 2016, MNRAS, 463, 2612

\bibitem[{Huang \& Chen(2025)}]{huang_physics-motivated_2025}
Huang, C. \& Chen, A.~Y. 2025, ApJ, 991, 90

\bibitem[{Johnston {et~al.}(1993)Johnston, Lorimer, Harrison, Bailes, Lynet,
  Bell, Kaspi, Manchester, D'Amico, Nleastrol, \&
  Shengzhen}]{johnston_discovery_1993}
Johnston, S., Lorimer, D.~R., Harrison, P.~A., {et~al.} 1993, Nature, 361, 613,
  aDS Bibcode: 1993Natur.361..613J

\bibitem[{Kalapotharakos {et~al.}(2021)Kalapotharakos, Wadiasingh, Harding, \&
  Kazanas}]{kalapotharakos_multipolar_2021}
Kalapotharakos, C., Wadiasingh, Z., Harding, A.~K., \& Kazanas, D. 2021, ApJ,
  907, 63

\bibitem[{Lockhart {et~al.}(2019)Lockhart, Gralla, Özel, \&
  Psaltis}]{lockhart_x-ray_2019}
Lockhart, W., Gralla, S.~E., Özel, F., \& Psaltis, D. 2019, MNRAS, 490, 1774

\bibitem[{Manchester \& Johnston(1995)}]{manchester_polarization_1995}
Manchester, R.~N. \& Johnston, S. 1995, The Astrophysical Journal, 441, L65,
  aDS Bibcode: 1995ApJ...441L..65M

\bibitem[{Mauviard {et~al.}(2025)Mauviard, Guillot, Salmi, Choudhury, Dorsman,
  González-Caniulef, Hoogkamer, Huppenkothen, Kazantsev, Kini, Olive,
  Stammler, Watts, Mendes, Rutherford, Schwenk, Svensson, Bogdanov, Kerr, Ray,
  Guillemot, Cognard, \& Theureau}]{mauviard_nicer_2025}
Mauviard, L., Guillot, S., Salmi, T., {et~al.} 2025, A {NICER} view of the 1.4
  solar-mass edge-on pulsar {PSR} {J0614}--3329, arXiv:2506.14883 [astro-ph]
  version: 1

\bibitem[{Miller {et~al.}(2019)Miller, Lamb, Dittmann, Bogdanov, Arzoumanian,
  Gendreau, Guillot, Harding, Ho, Lattimer, Ludlam, Mahmoodifar, Morsink, Ray,
  Strohmayer, Wood, Enoto, Foster, Okajima, Prigozhin, \&
  Soong}]{miller_psr_2019}
Miller, M.~C., Lamb, F.~K., Dittmann, A.~J., {et~al.} 2019, ApJL, 887, L24

\bibitem[{Miller {et~al.}(2021)Miller, Lamb, Dittmann, Bogdanov, Arzoumanian,
  Gendreau, Guillot, Ho, Lattimer, Loewenstein, Morsink, Ray, Wolff, Baker,
  Cazeau, Manthripragada, Markwardt, Okajima, Pollard, Cognard, Cromartie,
  Fonseca, Guillemot, Kerr, Parthasarathy, Pennucci, Ransom, \&
  Stairs}]{miller_radius_2021}
Miller, M.~C., Lamb, F.~K., Dittmann, A.~J., {et~al.} 2021, ApJ, 918, L28, aDS
  Bibcode: 2021ApJ...918L..28M

\bibitem[{Mitra {et~al.}(2023)Mitra, Melikidze, \& Basu}]{mitra_evidence_2023}
Mitra, D., Melikidze, G.~I., \& Basu, R. 2023, MNRAS: Letters, 521, L34

\bibitem[{Morsink {et~al.}(2007)Morsink, Leahy, Cadeau, \&
  Braga}]{morsink_oblate_2007}
Morsink, S.~M., Leahy, D.~A., Cadeau, C., \& Braga, J. 2007, ApJ, 663, 1244

\bibitem[{Navarro {et~al.}(1997)Navarro, Manchester, Sandhu, Kulkarni, \&
  Bailes}]{navarro_mean_1997}
Navarro, J., Manchester, R.~N., Sandhu, J.~S., Kulkarni, S.~R., \& Bailes, M.
  1997, ApJ, 486, 1019

\bibitem[{Osłowski {et~al.}(2014)Osłowski, van Straten, Bailes, Jameson, \&
  Hobbs}]{oslowski_timing_2014}
Osłowski, S., van Straten, W., Bailes, M., Jameson, A., \& Hobbs, G. 2014,
  MNRAS, 441, 3148

\bibitem[{Pétri(2017)}]{petri_polarized_2017}
Pétri, J. 2017, MNRAS, 466, L73

\bibitem[{Pétri {et~al.}(2023)Pétri, Guillot, Guillemot, Cognard, Theureau,
  Grießmeier, Bondonneau, González-Caniulef, Webb, Jankowski, Kravtsov,
  McKee, Carozzi, Cecconi, Serylak, \& Zarka}]{petri_constraining_2023-2}
Pétri, J., Guillot, S., Guillemot, L., {et~al.} 2023, A\&A, 680, A93

\bibitem[{Pétri {et~al.}(2025)Pétri, Guillot, Guillemot, González-Caniulef,
  Jankowski, Grießmeier, Theureau, \& Cognard}]{petri_double_2025-2}
Pétri, J., Guillot, S., Guillemot, L., {et~al.} 2025, A\&A, 701, A39

\bibitem[{Radhakrishnan \& Cooke(1969)}]{radhakrishnan_magnetic_1969}
Radhakrishnan, V. \& Cooke, D.~J. 1969, Ap. Lett., 3, 225

\bibitem[{Reardon {et~al.}(2024)Reardon, Bailes, Shannon, Flynn, Askew, Bhat,
  Chen, Curyło, Feng, Hobbs, Kapur, Kerr, Liu, Manchester, Mandow, Mishra,
  Russell, Shamohammadi, Zhang, \& Zic}]{reardon_neutron_2024}
Reardon, D.~J., Bailes, M., Shannon, R.~M., {et~al.} 2024, ApJL, 971, L18

\bibitem[{Riley {et~al.}(2019)Riley, Watts, Bogdanov, Ray, Ludlam, Guillot,
  Arzoumanian, Baker, Bilous, Chakrabarty, Gendreau, Harding, Ho, Lattimer,
  Morsink, \& Strohmayer}]{riley_nicer_2019}
Riley, T.~E., Watts, A.~L., Bogdanov, S., {et~al.} 2019, ApJL, 887, L21

\bibitem[{Salmi {et~al.}(2024)Salmi, Deneva, Ray, Watts, Choudhury, Kini,
  Vinciguerra, Cromartie, Wolff, Arzoumanian, Bogdanov, Gendreau, Guillot, Ho,
  Morsink, Cognard, Guillemot, Theureau, \& Kerr}]{salmi_nicer_2024}
Salmi, T., Deneva, J.~S., Ray, P.~S., {et~al.} 2024, ApJ, 976, 58

\bibitem[{Salmi {et~al.}(2022)Salmi, Vinciguerra, Choudhury, Riley, Watts,
  Remillard, Ray, Bogdanov, Guillot, Arzoumanian, Chirenti, Dittmann, Gendreau,
  Ho, Miller, Morsink, Wadiasingh, \& Wolff}]{salmi_radius_2022}
Salmi, T., Vinciguerra, S., Choudhury, D., {et~al.} 2022, ApJ, 941, 150, aDS
  Bibcode: 2022ApJ...941..150S

\bibitem[{Smith {et~al.}(2023)Smith, Abdollahi, Ajello, Bailes, Baldini,
  Ballet, Baring, Bassa, Gonzalez, Bellazzini, Berretta, Bhattacharyya,
  Bissaldi, Bonino, Bottacini, Bregeon, Bruel, Burgay, Burnett, Cameron,
  Camilo, Caputo, Caraveo, Cavazzuti, Chiaro, Ciprini, Clark, Cognard,
  Corongiu, Orestano, Crnogorcevic, Cuoco, Cutini, D’Ammando, Angelis,
  DeCesar, Gaetano, Menezes, Deneva, Palma, Lalla, Dirirsa, Venere, Domínguez,
  Dumora, Fegan, Ferrara, Fiori, Fleischhack, Flynn, Franckowiak, Freire,
  Fukazawa, Fusco, Galanti, Gammaldi, Gargano, Gasparrini, Giacchino,
  Giglietto, Giordano, Giroletti, Green, Grenier, Guillemot, Guiriec,
  Gustafsson, Harding, Hays, Hewitt, Horan, Hou, Jankowski, Johnson, Johnson,
  Johnston, Kataoka, Keith, Kerr, Kramer, Kuss, Latronico, Lee, Li, Li,
  Limyansky, Longo, Loparco, Lorusso, Lovellette, Lower, Lubrano, Lyne, Maan,
  Maldera, Manchester, Manfreda, Marelli, Martí-Devesa, Mazziotta, McEnery,
  Mereu, Michelson, Mickaliger, Mitthumsiri, Mizuno, Moiseev, Monzani,
  Morselli, Negro, Nemmen, Nieder, Nuss, Omodei, Orienti, Orlando, Ormes,
  Palatiello, Paneque, Panzarini, Parthasarathy, Persic, Pesce-Rollins,
  Pillera, Poon, Porter, Possenti, Principe, Rainò, Rando, Ransom, Ray,
  Razzano, Razzaque, Reimer, Reimer, Renault-Tinacci, Romani, Sánchez-Conde,
  Parkinson, Scotton, Serini, Sgrò, Shannon, Sharma, Shen, Siskind, Spandre,
  Spinelli, Stappers, Stephens, Suson, Tabassum, Tajima, Tak, Theureau,
  Thompson, Tibolla, Torres, Valverde, Venter, Wadiasingh, Wang, Wang, Wang,
  Weltevrede, Wood, Yan, Zaharijas, Zhang, \& Zhu}]{smith_third_2023}
Smith, D.~A., Abdollahi, S., Ajello, M., {et~al.} 2023, ApJ, 958, 191

\bibitem[{Sur {et~al.}(2024)Sur, Yuan, \& Philippov}]{sur_radio_2024}
Sur, A., Yuan, Y., \& Philippov, A. 2024, ApJ, 965, 140

\bibitem[{Vinciguerra {et~al.}(2024)Vinciguerra, Salmi, Watts, Choudhury,
  Riley, Ray, Bogdanov, Kini, Guillot, Chakrabarty, Ho, Huppenkothen, Morsink,
  Wadiasingh, \& Wolff}]{vinciguerra_updated_2024}
Vinciguerra, S., Salmi, T., Watts, A.~L., {et~al.} 2024, ApJ, 961, 62

\bibitem[{Zavlin(2006)}]{zavlin_xmm-newton_2006}
Zavlin, V.~E. 2006, ApJ, 638, 951, aDS Bibcode: 2006ApJ...638..951Z

\bibitem[{Zavlin {et~al.}(2002)Zavlin, Pavlov, Sanwal, Manchester, Trümper,
  Halpern, \& Becker}]{zavlin_x-radiation_2002}
Zavlin, V.~E., Pavlov, G.~G., Sanwal, D., {et~al.} 2002, ApJ, 569, 894, aDS
  Bibcode: 2002ApJ...569..894Z

\end{thebibliography}

\end{document}